\documentclass[%
 reprint,longbibliography,
superscriptaddress,
showpacs,preprintnumbers,
bibnotes,
 amsmath,amssymb,
 aps,
pra,
floatfix,
]{revtex4-2}
\usepackage{latexsym}
\usepackage{xcolor}
\usepackage{color}
\usepackage{graphicx}
\usepackage{bm}
\usepackage{times}
\renewcommand{\vec}[1]{\mathbf{#1}}
\newcommand{\ii}{\mathrm{i}}
\newcommand{\tn}[1]{\textnormal{#1}}
\renewcommand{\d}{\mathrm{d}}
\newcommand{\tc}{T_{\cal C}}
\newcommand{\dc}{\delta_{\cal C}}
\newcommand{\intc}{\int_{\cal C}}

\newcommand{\lowerbossphantom}{\vphantom{\bar{{x}}}}
\newcommand{\upperbossphantom}{\vphantom{\dagger}}
\newcommand{\Aop}[1]{\ensuremath{\chat{c}_{#1\lowerbossphantom}^{\upperbossphantom}}}
\newcommand{\Cop}[1]{\ensuremath{\chat{c}_{#1\lowerbossphantom}^{\dagger\upperbossphantom}}}
\newcommand{\xbar}[1]{{\mathchoice{\tempbar[\displaystyle]{#1}{3.5mu}}{\tempbar{#1}{3.5mu}}{\tempbar[\scriptstyle]{#1}{3.5mu}}{\tempbar[\scriptscriptstyle]{#1}{3mu}}}} 
\newcommand{\tempbar}[3][\textstyle]{\settowidth{\dimen1}{$#1\bar{#2}$}\makebox[\dimen1][l]{$#1\bar{#2\mspace{#3}}$}}
\newcommand{\cbar}[1]{\ensuremath{\xbar{#1}}}
\newcommand{\tempop}[3][\textstyle]{\settowidth{\dimen1}{$#1\hat{#2}$}\makebox[\dimen1][l]{$#1\hat{#2\mspace{#3}}$}}
\newcommand{\xop}[1]{{\mathchoice{\tempop[\displaystyle]{#1}{3.5mu}}{\tempop{#1}{3.5mu}}{\tempop[\scriptstyle]{#1}{3.5mu}}{\tempop[\scriptscriptstyle]{#1}{3mu}}}}
\newcommand{\chat}[1]{\ensuremath{\xop{#1}}}

\begin{document}
\title{Doublon production in correlated materials by multiple ion impacts}
\author{Lotte Borkowski}
\author{Niclas Schl\"{u}nzen}
\author{Jan Philip Joost}
\author{Franziska Reiser}
\author{Michael Bonitz}
\affiliation{Institut f\"{u}r Theoretische Physik und Astrophysik, Christian-Albrechts-Universit\"{a}t zu Kiel, Leibnizstrasse 15, 24098, Kiel, Germany}

\begin{abstract}
In a recent Letter [Balzer \textit{et al.}, Phys. Rev. Lett. \textbf{121}, 267602 (2018)] it was demonstrated that ions impacting a correlated graphene cluster can excite strongly nonequilibrium states. In particular, this can lead to an enhanced population of bound pairs of electrons with opposite spin --- doublons --- where the doublon number can be increased via multiple ion impacts. These predictions were made based on nonequilibrium Green functions (NEGF) simulations allowing for a time-dependent non-perturbative study of the energy loss of  charged particles penetrating
a strongly correlated system. Here we extend these simulations to larger clusters and longer simulation times, utilizing the recently developed G1--G2 scheme [Schünzen \textit{et al.}, Phys. Rev. Lett. \textbf{124}, 076601 (2020)] which allows for a dramatic speedup of NEGF simulations. Furthermore, we investigate the dependence of the energy and doublon number on the time interval between ion impacts and on the impact point.
\end{abstract}
\pacs{05.30.-d, 34.10.+x, 34.50.Bw, 71.10.Fd}
\maketitle

\section{\label{sec.introduction}Introduction}

The interaction of charged particles with matter is 
of high relevance for many areas of physics and astrophysics. Processes such as the stopping power or stopping range of projectiles in matter allow one to analyze their energy spectrum. On the other hand, the energy loss of charged particles in matter is a sensitive tool to diagnose the electronic properties of the material.
In soft collisions of heavy charged particles,
such as ions, with a solid, typically the electrostatic force, i.e., the Coulomb interaction,
has the largest impact, leading to excitation and ionization of electrons in the target material
and, thus, to the loss of kinetic energy of the projectile~\cite{sigmund06}. For nonrelativistic
projectile velocities of the order of or larger than the Fermi velocity ($\sim10^6$\,m/s
in metals), theoretical approaches based on scattering theory~\cite{nagy98} or on the linear response
functions of the uniform electron gas~\cite{pitarke95} provide an accurate  description
of the energy transferred during the collision process. However, they neglect the precise atomic composition
of the target, electronic correlations, as well as nonlinear effects  \cite{echenique_pra_86, dornheim_prl_20}.

In the same velocity regime, recent theoretical progress
is due to time-dependent density functional theory (TDDFT), which has been applied to describe
the slowing down of charged particles in a variety of solids, including metals~\cite{quijada07,zeb12,schleife15},
semimetals~\cite{ojanperae14,Zhao_2014} and clusters~\cite{bubin12,mao14}, narrow-band-gap
semiconductors~\cite{ullah15}, insulators~\cite{pruneda07,zeb13} and monolayer systems such as graphene, e.g. Ref.~\cite{konova_acs.nanolett_21}. Taking into account
primarily the excitation of valence electrons, these simulations yield satisfactory results
for the electronic stopping power (the transfer of energy to the electronic degrees of freedom
per unit length traveled by the projectile) and work for a wide range of impact energies. On
the other hand, a rather general tool to determine the stopping power of energetic ions in matter
is provided by the SRIM code~\cite{ziegler_srim_2010}, which uses the binary collision approximation in combination
with an averaging over a large range of experimental situations, and data being available for many materials and gaseous targets. At the same time, linear response,  TDDFT and SRIM have difficulties in accounting for strong  electron-electron correlations which are crucial, e.g., in transition
metal oxides~\cite{anisimov97} or specific organic materials~\cite{singla2015}. 
In addition, we
note that SRIM and linear response theory do not account 
for time-dependent changes in the target during the collision process, which limits their
applicability.

For this reason we have developed an alternative approach to the stopping power that is base on nonequilibrium Green functions~(NEGF)~\cite{kadanoff-baym-book,stefanucci13_cambridge,balzer-book,schluenzen_jpcm_19}.
This method allows one to systematically include electron-electron correlations via a time-dependent
many-body selfenergy, and it has recently successfully been applied to strongly correlated lattice systems as well~\cite{schluenzen_prb16,schluenzen_prb17}. Particular advantages of the NEGF approach are that it is not
limited to either weak or strong coupling and that it is particularly well suited to study finite-sized
clusters and spatially inhomogeneous systems on a self-consistent footing. Recently, NEGF simulations coupled to an Ehrenfest dynamics of the projectile were developed \cite{balzer_prb16}, and good agreement with TDDFT and SRIM simulations was established. A particular interesting result of  NEGF simulations in the low velocity range was the prediction of a non-trivial electronic correlation effect: the formation of doublons, i.e. bound pairs of electrons with opposite spin, as a results of ion impact. In a recent Letter~\cite{balzer_prl_18} it was demonstrated that the doublon number can be further increased if the material is hit by multiple ions. This issue was further explored in Refs.~\cite{bonitz_pss_18,schluenzen_cpp_18}. Another important issue during the impact of ions on solid targets is the possible transfer of charge to the ion (neutralization) which has been studied in a number of experiments, e.g. Refs.~\cite{Aumayr_2008,Gruber2016}. A first attempt to extend the NEGF-Ehrenfest approach to include charge exchange processes has been made in Ref.~\cite{balzer_cpp_21}.

On the other hand, the NEGF approach
is computationally very demanding. While in recent years efficient numerical schemes have been developed
to solve the underlying Keldysh--Kadanoff--Baym equations (KBE)~\cite{dahlen07,stan09,balzer_pra_10,balzer_pra_10_2,garny10,balzer-book,latini14,hermanns_prb14}, 
the solution is hampered by a cubic scaling of the CPU time with the number of time steps $N_{\rm t}$. This scaling can be reduced to quadratic by applying the generalized Kadanoff-Baym ansatz (HF-GKBA). Recently we could show that even linear scaling can be achieved if the HF-GKBA is rewritten as a coupled system of time-local equations for the one- and two-particle Green function which was called ``G1--G2 scheme''~\cite{schluenzen_20_prl}.

In the present paper we take advantage of the speedup provided by the G1--G2 scheme to significantly extend the previous
stopping simulations for multiple ion impacts of Ref.~\cite{balzer_prl_18}. We study significantly larger hexagonal monolayer clusters of up to $96$ sites, extend the simulation duration, and study how the excitation by the ion propagates through the cluster. 
Further,
we vary the time interval between successive ion impacts as well as the impact point and investigate how this influences the energy of the electrons and the doublon production. For short time intervals between impacts we are able to analyze non-adiabatic response effects where the effect of two projectiles does not simply add up because the cluster is driven far away from equilibrium. In addition to the long-range Coulomb interaction between cluster electrons and projectile we also consider a simpler model where the effect of the projectile is mimicked by a variation of the lattice potential. This scenario is of interest as well as it is easily realized for cold atoms in an optical lattice.

The paper is organized as follows. In Sec.~\ref{sec.method}, we define the model Hamiltonian and introduce our NEGF approach as well as the G1--G2 scheme.
In
Sec.~\ref{sec.results.stopping.all}.
 we present the simulation results. We  conclude the paper with Sec.~\ref{sec.discussion},
outlining possible future work.


\section{\label{sec.method}Theory}

\subsection{\label{subsec.model}Hubbard model for finite graphene clusters}

To study the stopping dynamics of a classical charged particle which passes through a
 correlated system, we consider a finite lattice of electrons described by a
single-band Fermi--Hubbard model and compute the energy exchange  between projectile and cluster during the collision
process. Details of the model are described in Ref.~\cite{balzer_prb16}. Here we only summarize the main points. The lattice as a whole is electrically neutral, i.e., the electronic
charges are compensated by corresponding opposite charges located at the site coordinates
$\vec{R}_i$. 
The energy exchange occurs via  the  Coulomb potential
between the projectile, the fixed background charges and the target electrons which are
initially in equilibrium.
As the projectile we consider positively charged ions.
When they approach the lattice they induce a confinement potential to the electrons which
 initiates a nonequilibrium electron dynamics.  

\begin{figure}[t]
 \includegraphics[width=0.403\textwidth]{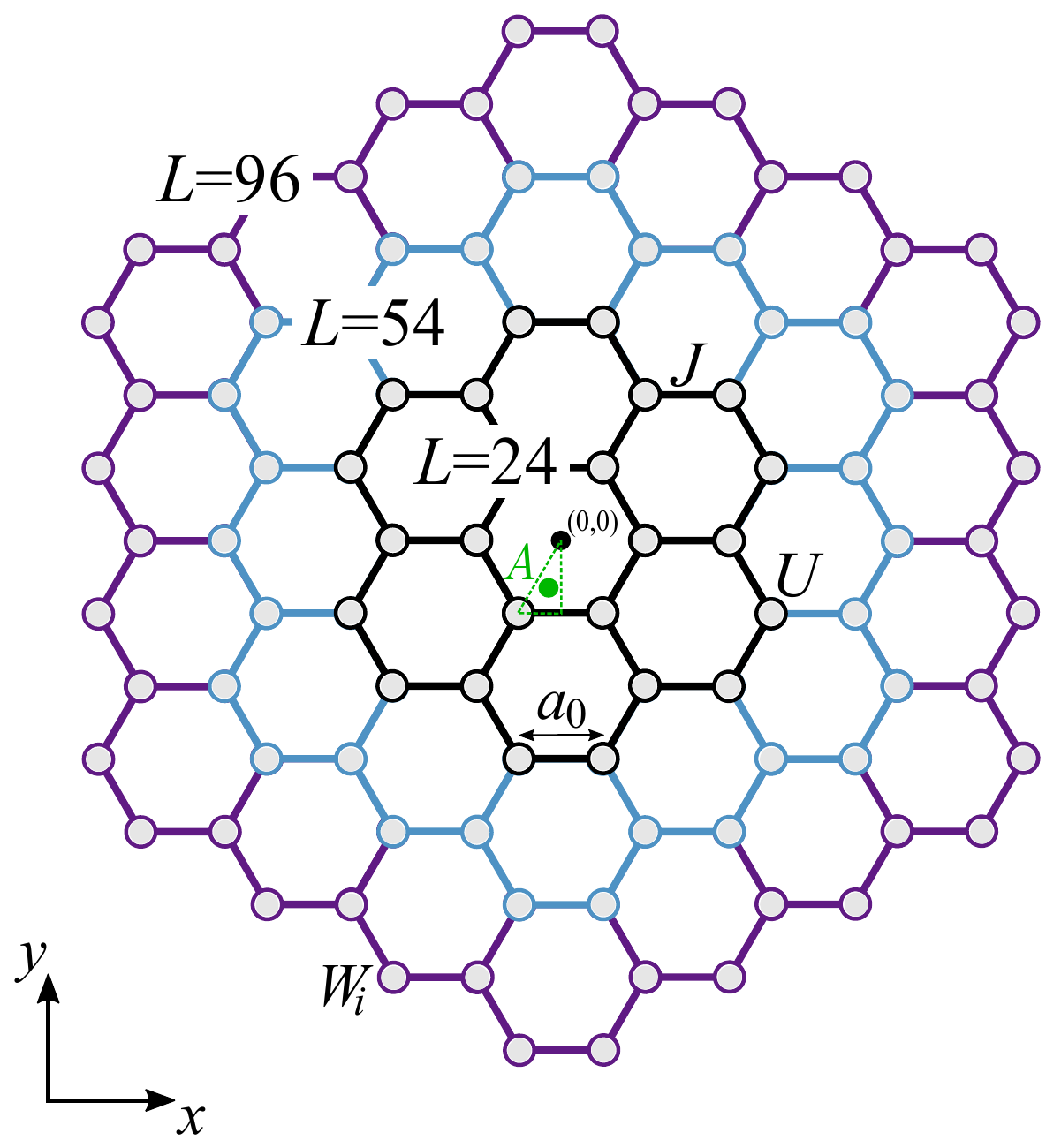}
 \caption{
 Lattice structure of circular honeycomb clusters with $L=24$
 (black), $54$ (blue) and 96 (purple) sites. The green point $A=\left(-\tfrac{1}{6}a_0,-\tfrac{\sqrt{3}}{3}a_0,0\right)$ indicates the position where the projectile
 hits the lattice plane. 
 Furthermore, $a_0=1.42$\AA $\,$ denotes the lattice spacing, $J=2.8$eV  is the nearest-neighbor hopping, $U$
 the on-site interaction, and $W_{i}$ is the time-dependent local energy produced by the projectile, cf. Eq.~(\ref{eq.ham2}).}
 \label{Fig.lattice}
\end{figure}

As a target, we consider circular honeycomb clusters  in the
$xy$-plane with a finite number of honeycombs and a total of $L$ sites, see
Fig.~\ref{Fig.lattice} for an illustration. We focus on a half-filled system in the
paramagnetic phase and use a lattice spacing of $a_0=1.42$\,{\AA},
which corresponds to the carbon-carbon bond length in graphene~\cite{katsnelson12}.
Using a nearest neighbor-hopping $J$ and an on-site Hubbard repulsion $U$ (which we consider time-dependent to account for adiabatic-switching procedures~\cite{schluenzen_jpcm_19}), the Hamiltonian
for the lattice electrons is then given by
\begin{align}
\label{eq.ham1}
 H_\textup{e}(t)&=-J\sum_{\langle i,j\rangle,\sigma} \Cop{i\sigma} \Aop{j\sigma}+U(t)\sum_{i}\chat{n}_{i}^\uparrow\chat{n}_{i}^\downarrow&\nonumber\\
&\hspace{1pc}+\sum_{i,\sigma}W_{i}(t) \Cop{i\sigma} \Aop{i\sigma}\,,
\end{align}
where the operator $\Cop{i\sigma}$ ($\Aop{i\sigma}$) creates (annihilates) an electron
with spin $\sigma$ on site $i$. Further,  $\chat{n}_{i\sigma}=\Cop{i\sigma} \Aop{i\sigma}$ and $W_{i}$ denote the
electron density operator and the external potential produced by the projectile, respectively.
%

For convenience, we measure $J$ and $U$ in electron volts, define $U/J$ as the
interaction strength for the electrons and use $t_0=\hbar/J$ as the unit of time.
Unless otherwise stated, we use $J=2.8$\,eV which is typical for graphene~\cite{Schueler2013} and corresponds to $t_0=0.235$fs.

\subsection{\label{subsec.projectile}Incident-ion potential}
We consider two different models to describe the time-dependent potential induced by the energetic ions. In the first scenario, we compute the classical \textbf{Coulomb} potential of incident ions of charge $Z_p$, 
\begin{align}
\label{eq.ham2}
 W_{i}(t)=-\frac{e^2}{4\pi\epsilon_0}\frac{Z_\textup{p}}{|\vec{r}_\textup{p}(t)-\vec{R}_i|}\,,
\end{align}
where $\vec{r}_\textup{p}(t)$ denotes the time-dependent position of the projectile, $-e$ is
the electron charge and $\epsilon_0$ the vacuum permittivity. 
In analogy to previous studies~\cite{balzer_prb16,schluenzen_cpp_18}, we set the inital position of the incident ion to $\vec{r}_{\textup{p}}(0)=\left(-\tfrac{1}{6}a_0,-\tfrac{\sqrt{3}}{3}a_0,-z\right)$,
see the centroid point of the green dashed triangle in Fig.~\ref{Fig.lattice}. These coordinates
have been found to give similar stopping results for the highly symmetric honeycomb lattice
compared to calculations, where one averages over many different collision sites.
The influence of the impact point will be studied separately in Sec.~\ref{ss:impact-point}.
Furthermore, the initial $z$-position is chosen such that the measured energy transfer becomes
independent of the initial conditions (typically $z\gtrsim10a_0$).
The dependence of the energy exchange on the impact velocity was investigated in detail in Refs.~\cite{balzer_prb16,balzer_prl_18}. Here we are interested in the response of the target electrons to a varying number of projectiles. To this end we fix the ion velocity to a value $v_{p,0}=3a_0/t_0\approx 1.8\cdot 10^6 m/s$ in $z$-direction, cf. Eq.~(\ref{eqn.yukawaCoulombInput}). This is a comparatively large value close to the maximum of the stopping curve. The corresponding kinetic energy of the projectile is much larger than the energy exchanged with the target. Therefore, a selfconsistent solution of Newton's equation for the projectile can be avoided for all cases considered in this paper.

It is often useful to simplify the general Coulomb potential to investigate local effects induced by the excitation. For such applications, a short-ranged localized potential that excites a specific lattice site provides a reasonable approximation \cite{balzer_prl_18}. We will use a \textbf{Gaussian},
\begin{align}
    W_i(t) = -W_0 \delta_{i,i_0} \exp{\left(-\frac{t^2}{2\tau^2}\right)} \, ,
    \label{eq:gaussian}
\end{align}
with the amplitude $W_0$, the excited lattice site $i_0$ (which we choose on the innermost honeycomb ring), and the interaction duration $\tau > 0$. This model is closely related to the full Coulomb model where $\tau$ is inversely proportional to the ion velocity, while $W_0$ is proportional to the  charge of the ion~\cite{balzer_prl_18}. For the best correspondence between the two models we use $\tau = 0.5\hbar/J $ and $W_0=8J$. We note that the Gaussian potential is also of practical relevance for experiments with ultracold atoms in optical lattices and allows to simulate ion stopping (for a recent overview, see Ref.~\cite{gross_science_2017}).

\subsection{\label{subsec.method}Nonequilibrium Green functions}

We compute the correlated time evolution of the lattice electrons, using a nonequilibrium Green functions (NEGF) approach, as described in Ref.~\cite{balzer_prb16}. The central quantity is the one-particle Green function [here and in the following we only give the spin-up components ($\uparrow$) explicitly, the spin-down components follow from replacing $\uparrow\, \leftrightarrow \,\downarrow$]
\begin{align}
\label{eq.negf}
 G_{ij}^\uparrow(t,t')=-\frac{\ii}{\hbar}\langle \tc \Aop{i\uparrow}(t) \Cop{j\uparrow}(t')\rangle\,,
\end{align}
which is defined as an ensemble average on the Keldysh time contour $\cal C$~\cite{keldysh64},
and $\tc$ 
denotes the contour time-ordering operator.
The equations of motion of the greater and less components of the NEGF (\ref{eq.negf}),
\begin{align}
 G^{>,\uparrow}_{ij}(t,t')&=-\frac{\ii}{\hbar}\langle \Aop{i\uparrow}(t)\Cop{j\uparrow}(t')\rangle\,,\\
 G^{<,\uparrow}_{ij}(t,t')&=\frac{\ii}{\hbar}\langle \Cop{j\uparrow}(t')\Aop{i\uparrow}(t)\rangle\,,\nonumber
\end{align}
are the two-time Keldysh--Kadanoff--Baym
equation (KBE)~\cite{kadanoff-baym-book,stan09,stefanucci13_cambridge}:
\begin{align}
\label{eq.kbe}
 \sum_k&[\ii\hbar\,\partial_t\delta_{ik}-h^{\tn{HF},\uparrow}_{ik}(t)]G_{kj\sigma}^{\gtrless,\uparrow}(t,t')&\\
 &=\dc(t,t')\delta_{ij}+\sum_{k}\left\{\intc \d s\,\Sigma^\uparrow_{ik}(t,s)G^\uparrow_{kj}(s,t')\right\}^\gtrless\,.\nonumber
\end{align}
Here, $\dc$ denotes the delta function on the contour, and $h^{\tn{HF},\uparrow}_{ij}(t)$ is the
time-dependent effective one-particle Hamiltonian, which explicitly includes the
Hartree contribution to the electron-electron interaction,
\begin{align}
\label{eq.hij}
h^{\tn{HF},\uparrow}_{ij}(t)=-\underbrace{J\delta_{\langle i,j\rangle}}_{=J_{ij}}+\left[W_{i}(t)-\ii\hbar U(t)G_{ii}^{<,\downarrow}(t)\right]\delta_{ij}\,.
\end{align}
On the right-hand side of Eq.~(\ref{eq.kbe}), the contour integral defines the memory
kernel of the KBE, in which $\Sigma^\uparrow_{ij}(t,t')$ denotes the correlation part
of the selfenergy [i.e., the mean-field part is excluded as it is contained in
Eq.~(\ref{eq.hij})]. Systematic expressions for the selfenergy can be constructed
by many-body perturbation theory, e.g.,  using diagram techniques~\cite{stefanucci13_cambridge,schluenzen_cpp16}.
Below, we treat the correlation selfenergy $\Sigma$ in two approximations,
which conserve particle number, momentum and energy.

%
\begin{enumerate}
\item Mean field approximation [time-dependent Hartree(-Fock)], here correlation effects are neglected.
\item Second-order Born approximation (SOA),
 \begin{align}
 \label{eq.sigma.2b}
  \Sigma_{ij}^{\lessgtr,\uparrow}(t,t')=&-\left(\ii\hbar\right)^2U(t)U(t') \nonumber \\
  &\times G^{\lessgtr,\uparrow}_{ij}(t,t') G^{\lessgtr,\downarrow}_{ij}(t,t') G^{\gtrless,\downarrow}_{ji}(t',t)\,,
 \end{align}
which includes all irreducible diagrams of second order in the interaction $U$.
We note that the SOA selfenergy is a perturbation theory result and, therefore,
becomes less accurate when $U$ increases. In the context of ion-impact simulations on finite graphene clusters, the accuracy of the SOA scheme has been validated in Ref.~\cite{balzer_prl_18}.
\end{enumerate}
A detailed recent overview on different selfenergy approximations and their accuracy can be found in the review by Schlünzen \textit{et al.}
\cite{schluenzen_jpcm_19}.

\subsection{G1--G2 scheme}\label{ss:g1-g2}
The advantage of NEGF simulations is that they allow to accurately capture electronic correlation effects. At the same time, NEGF simulations are computationally expensive because the CPU time scales cubically with the number of time steps,  $\sim N^3_{\rm t}$.
In the recent publications, Refs.~\cite{schluenzen_20_prl} and \cite{joost_prb_20}, it was demonstrated that the KBE in combination with the HF-GKBA~\cite{lipavsky86,hermanns_prb14} can be solved highly efficiently within a set of time-local differential equations, bringing the CPU time scaling down to $N^1_{\rm t}$. In this description, the equation of motion for the time-diagonal single-particle Green function becomes~\cite{joost_prb_20}  
\begin{align}
\ii\hbar \frac{\d}{\d t} G^{<,\uparrow}_{ij}(t) &= \left[h^{\tn{HF},\uparrow}(t), G^{<,\uparrow}(t)\right]_{ij} + \left[I+I^\dagger\right]^{\uparrow}_{ij}(t) \, . \label{eq:eom_g}
\end{align}
The information of the selfenergy is included in the correlation part of time-diagonal two-particle Green function $\mathcal{G}$ as
\begin{align}
 I^{\uparrow}_{ij}(t) &= \sum_k \int_{t_0}^t \d\cbar t\, \left[ \Sigma^{>,\uparrow}_{ik}(t,\cbar t) G^{<,\uparrow}_{kj}(\cbar t,t)\right. \nonumber \\&\qquad\qquad\qquad\left.- \Sigma^{<,\uparrow}_{ik}(t,\cbar t) G^{>,\uparrow}_{kj}(\cbar t,t) \right] \nonumber \\
 &= -\ii\hbar U(t) \mathcal{G}^{\uparrow\downarrow\uparrow\downarrow}_{iiji}(t)\, . \label{eq:colint_G2}
\end{align}
The correlation function $\mathcal{G}$ obeys its own equation of motion, which, for the SOA selfenergy [cf. Eq.~\eqref{eq.sigma.2b}], attains the form
\begin{align}
 \ii\hbar\frac{\d}{\d t} \mathcal{G}^{\uparrow\downarrow\uparrow\downarrow}_{ijkl} (t) - \left[h^{(2),\tn{HF}}_{\uparrow\downarrow}, \mathcal{G}^{\uparrow\downarrow\uparrow\downarrow}\right]_{ijkl} (t) =& \Psi^{\uparrow\downarrow\uparrow\downarrow}_{ijkl}(t)\, , \label{eq:eom_g2}
\end{align}
with the two-particle Hartree--Fock (HF) Hamiltonian
\begin{align}
 h^{(2),\tn{HF}}_{ijkl,\uparrow\downarrow}(t) = \delta_{jl} h^{\tn{HF},\uparrow}_{ik}(t) + \delta_{ik}h^{\tn{HF},\downarrow}_{jl}(t)\, , \label{eq:twopart_hamiltonian_hubbard}
\end{align}
and the two-particle source term
\begin{align}
\Psi^{\uparrow\downarrow\uparrow\downarrow}_{ijkl}(t) &= \left(\ii\hbar\right)^2 U(t) \sum_{p}
  \left[ G^{>,\uparrow}_{ip}(t) G^{>,\downarrow}_{jp}(t) G^{<,\uparrow}_{pk}(t) G^{<,\downarrow}_{pl}(t)\right.\nonumber \\
  &\left.\qquad\qquad\qquad-G^{<,\uparrow}_{ip}(t) G^{<,\downarrow}_{jp}(t) G^{>,\uparrow}_{pk}(t) G^{>,\downarrow}_{pl}(t) \right] \, .
\label{eq:phi-hubbard}  
\end{align}
The coupled equations~\eqref{eq:eom_g} and \eqref{eq:eom_g2} are called the G1--G2 scheme and are closely related to the density-operator formalism~\cite{bonitz_qkt}. For this work, the equations are propagated with a fourth-order Runge--Kutta integration scheme. 
From the resulting $G^<$ and $\mathcal{G}$, we have direct access to the observables of interest, namely the site-resolved densities and double occupation
\begin{align}
n_{ij}(t) &= n^\uparrow_{ij}(t) + n^\downarrow_{ij}(t) =-\ii\hbar\left[G^{<,\uparrow}_{ij}(t) + G^{<,\downarrow}_{ij}(t)\right]\, , \\
d_i(t) &= n^\uparrow_{ii}(t) n^\downarrow_{ii}(t) + \left(\ii\hbar\right)^2 \mathcal{G}_{iiii}^{\uparrow\downarrow\uparrow\downarrow}(t) \, ,
\label{eq:d-def}
\end{align}
where $d_i$ consists of a mean field (first term) and a correlation contribution (second term). Below we will also compute the cluster-averaged doublon number
\begin{align}
    d_{\tn{av}}(t) = \frac{1}{L} \sum_{i} d_i(t) \, .
    \label{eq:dav-def}
\end{align}
Additional important observables are the total energy of the target, 
\begin{align}
    E(t) = E_\tn{kin}(t) + E_\tn{pot}(t) + E_\tn{int}(t)\, ,
    \label{eq:etot-def}
\end{align}
comprising the kinetic, potential and interaction energy contributions:
\begin{align}
    E_\tn{kin}(t) &= - J \sum_{ij} \delta_{\left<i,j\right>} n_{ji}(t)\, , \\
    E_\tn{pot}(t) &=  \sum_{i} W_{i}(t) n_{ii}(t) \, , \\
    E_\tn{int}(t) &= U(t) \sum_{i} d_i(t)\, . \label{eq:kinetic_energy}
\end{align}

\section{\label{sec.results.stopping.all}Numerical results}

\subsection{\label{sec.results.stopping1}Benchmarks against previous HF-GKBA simulations}
\begin{figure}[h]
\centering
	\includegraphics[scale=1]{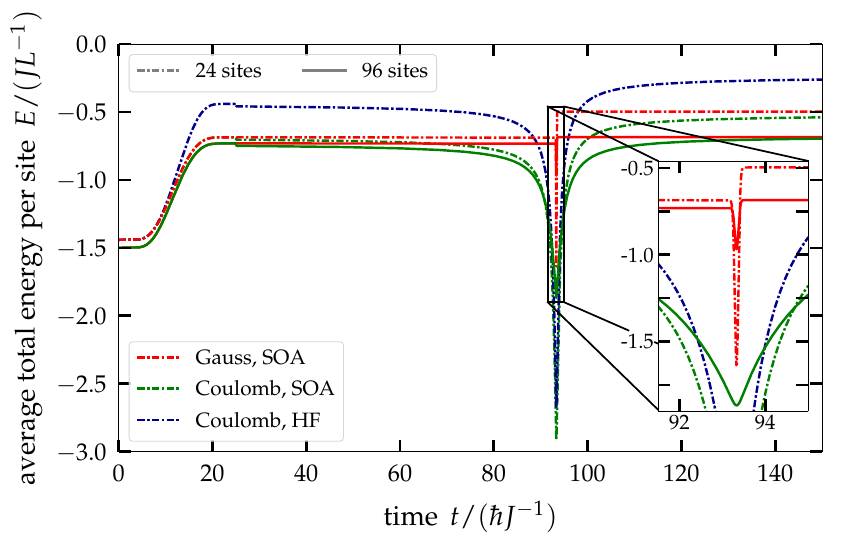}
\caption[Single impact energy dynamics for different potentials, self energies and lattice sizes]{Average total energy per site of two different lattice sizes, $L = 24$ (dash-dotted lines) and $L = 96$ (solid lines) for a single excitation of the system around $t = 93 \, t_0$ with $t_0\approx 0.235$fs. The three different colors denote different potentials and self energies. For the blue and green curves the Coulomb potential is used to mimic the projectile impact with either time-dependent Hartree-Fock (HF, blue) or second-order Born (SOA, green) as selfenergy approximation. The red curves correspond to an impact mimicked by the Gaussian potential in second-order Born approximation.}
\label{Fig.singleimpact}
\end{figure}

When an ion penetrates through the target, the electric potential forces the lattice electrons to accumulate near the impact point. For sufficiently fast projectiles this corresponds to a non-adiabatic excitation, leading to an overall energy gain for the cluster electrons. 
This behavior is shown in Fig.~\ref{Fig.singleimpact}, which gives a general overview of the impact dynamics. The figure 
depicts the total time evolution of the total energy of the target averaged over the number of lattice sites for the case of a single projectile impact. We compare two finite hexagonal systems containing 24 and 96 sites, respectively. Further, we study the influence of the projectile models presented in Sec.~\ref{subsec.projectile}.
Finally, the effect of the electron--electron interaction in the target is analyzed by comparing results for Hartree--Fock (HF) and second-order Born (SOA) selfenergies.

Let us briefly summarize what we observe in Fig.~\ref{Fig.singleimpact}. First, one clearly distinguishes four phases. The fist, from $t=0$ to $\approx 25 t_0$ depicts the adiabatic switch-on of correlations, which allows us to start the simulations from an uncorrelated initial state. Thus, around $t=25t_0$, a correlated initial state is achieved which is crucial for a selfconsistent dynamical treatment of the projectile--target interaction. Note that the small kink in the energy, at $t=25 t_0$, is due to the switch on of the electron--ion interaction which is very small but finite for the chosen initial position of the ion. The second and third phases describe the ion approaching and penetrating through the graphene-flake around $t=93 t_0$. Finally, the fourth phase describes the departure of the ion after traversal of the target. The increased total energy of the lattice corresponds to the projectile's energy loss, which we cannot measure directly within our description.

The amount of transferred energy depends on the electronic correlations in the target and on the theoretical model: the energy gain of the target is higher for HF results than for SOA, in agreement with earlier studies \cite{balzer_prb16}. If the lattice size is increased, the transferred energy is distributed among a larger number of lattice electrons. As a consequence, the energy gain per lattice site is reduced (compare the dashed and full red lines in the inset).
%

Regarding the models of the ionic potential, we observe that---with a suitable choice of parameters---the Gaussian model reproduces the energy gain of the lattice very well (compare the initial and final energies of the red and green curves in phases 2 and 4). However, there are drastic differences during the impact phase. Here the full long-range Coulomb model predicts significantly larger intermediate energy exchange between projectile and target electrons. This is due to the long range of the Coulomb potential, that affects a larger number of lattice charges. 
This effect is even more pronounced in the case of multiple ion impacts (see Sec.~\ref{subsec.GaussCoulomb}). Finally, the increased magnitude of the energy exchange is also important for a proper analysis of the details of the excitation mechanism which involves two-electron excitations and doublon formation \cite{balzer_prl_18,bonitz_pss_18}. 


\subsection{Multiple ion impacts. Gauss model vs. Coulomb interaction}
\label{subsec.GaussCoulomb}
\begin{figure}[h]
\centering
	\includegraphics[scale=1]{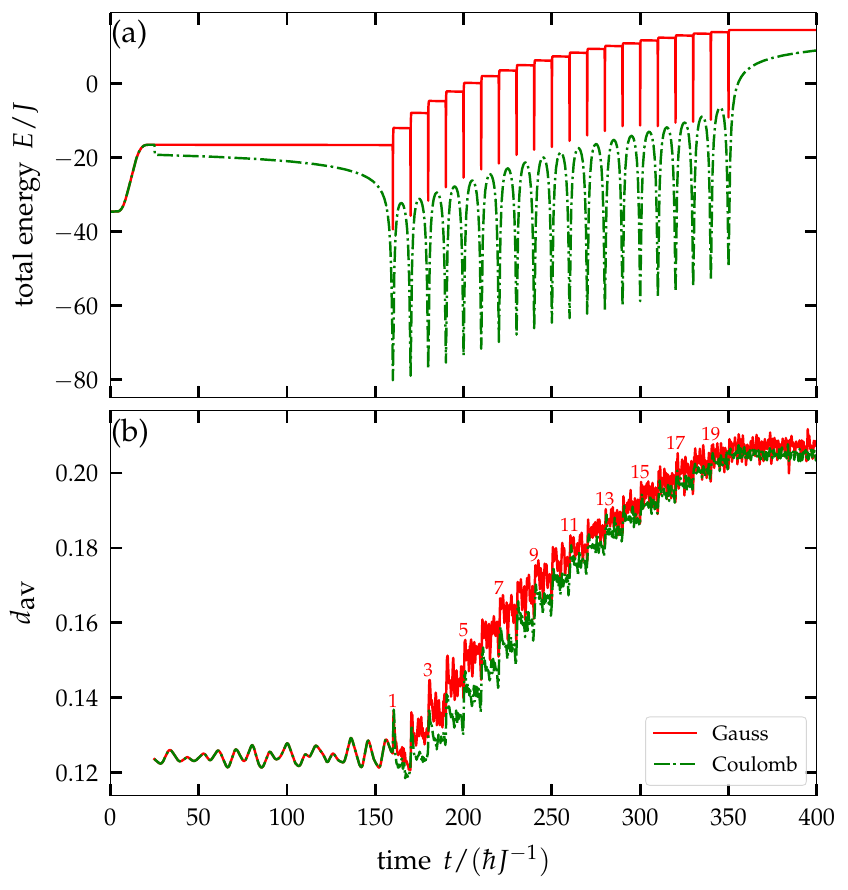}
\caption[Total energy and double occupation dynamics for different potentials and $L = 24$]{Top: total energy  increase of the lattice containing $L = 24$ sites (cf.~Fig.~\ref{Fig.lattice}) induced by 20 projectile impacts with charge $Z_{\textnormal{p}} e = e$. Comparison of the Gaussian model (red) and the  Coulomb interaction potential (green).
(b) Cluster averaged doublon number. Numbers in the figure denote the number of the impacting ion.
The model details are described in Sec.~\ref{subsec.projectile}.
}
\label{Fig.potentialEnergyDoubOcc}
\end{figure}

When increasing the number of projectiles impacting the honeycomb cluster, a general total energy gain in the lattice can be observed as shown in Fig.~\ref{Fig.potentialEnergyDoubOcc}~(a).
It depicts the total energy dynamics of the smallest lattice introduced in Fig.~\ref{Fig.lattice} containing $L = 24$ sites.
In total, this lattice was excited 20 times ($N_\tn{x} = 20$) for the Gaussian (red) and the Coulomb (green) potential respectively.
Thereby, every peak corresponds to a projectile impact, while the different depths of the peaks can be attributed to the disparate shape and range of the potentials.
Both curves describe an ion with charge $Z_p=1$ with an initial distance and velocity of
\begin{align}
\label{eqn.yukawaCoulombInput}
	\vec{r}_{\textnormal{p},0} = -\left(\begin{array}{c} \frac{1}{6} \\[1ex] \frac{\sqrt{3}}{3} \\[1ex] 300 \end{array}\right)a_0, \qquad
	\vec{v}_{\textnormal{p},0} = \left(\begin{array}{c} 0 \\ 0 \\ 3 \end{array}\right) \frac{a_0}{t_0}\, ,
\end{align}
with $t_0=\frac{\hbar}{J}\approx 0.235$fs and $v_{p,0}\approx 1.813\cdot 10^6$m/s.\\
The coordinate $\textbf{r}_{p,0}$, corresponds to the green point $A$ in Fig.~\ref{Fig.lattice} and was used as impact point for the Coulomb potential unless stated otherwise. In contrast, the Gaussian potential is directly applied to one of the innermost lattice sites.\\
Figure~\ref{Fig.potentialEnergyDoubOcc} (a) shows an initial increase of the total energy until $t = 25 \, t_0$, which is due to the adiabatic switch-on (AS) of electronic correlations in the solid \cite{schluenzen_jpcm_19}.
After the AS, the respective time-dependent interaction potential with the projectile is turned on. 
The small kink in the green curve for the total energy at $t = 25 \, t_0$ is due to the long range of the Coulomb interaction. This kink could be easily eliminated by a different procedure of turning on the projectile-electron interaction, but this would require additional computational cost. 
When the projectile approaches the cluster plane the total electron energy rapidly decreases and increases again, when the ion leaves. With each new ion this behavior occurs again, in agreement with Fig.~\ref{Fig.singleimpact}.
Next, we focus on the time evolution of the double occupation, Eq.~(\ref{eq:dav-def}), which is plotted in Fig.~\ref{Fig.potentialEnergyDoubOcc} (b).
We observe that, through multiple periodical excitations, the overall double occupation in the two-dimensional finite cluster can be significantly increased regardless of the potential chosen to mimic the projectile.
In the figure, every second excitation is numbered. The instant when an ion passes through the lattice plane is clearly visible from the peaks of $d_{\textnormal{av}}$.
If the number of impacts would be increased further, both curves are expected to eventually reach $d_{\textnormal{av}} = 0.25$ -- the value in an uncorrelated system at half filling, cf. Eq.~(\ref{eq:d-def}). This is also confirmed by the time evolution of the correlation energy, which steadily decreases with increasing number of excitations.

\begin{figure}[h]
\centering
	\includegraphics[scale=1]{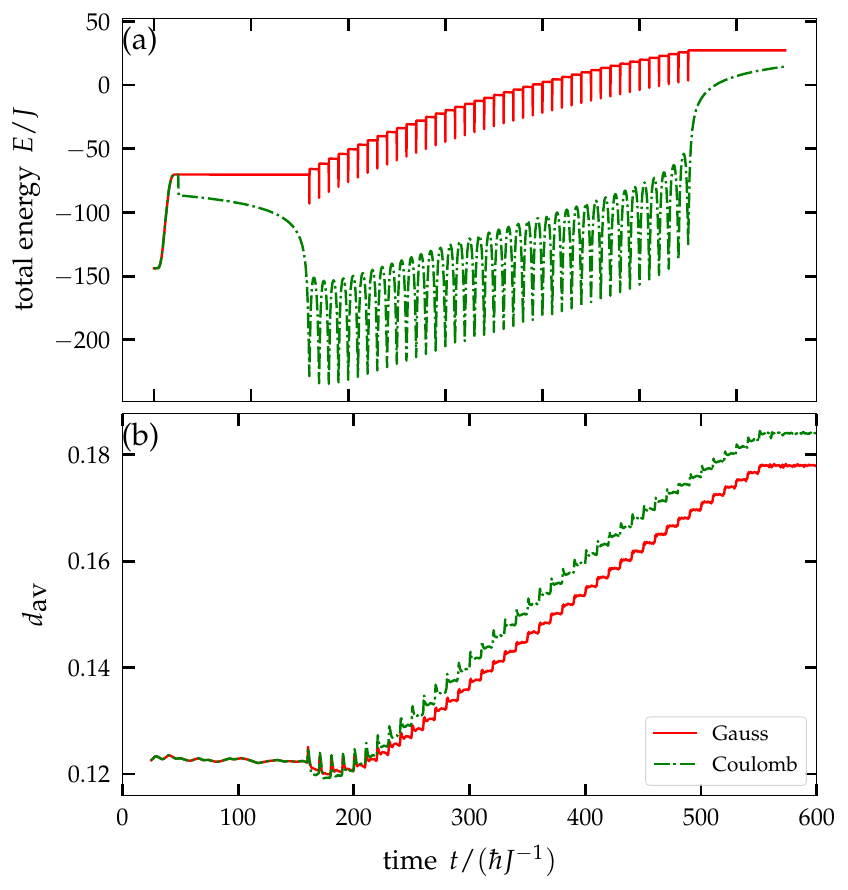}
\caption[Total energy and average double occupation dynamics for different potentials and $L = 96$]{Total energy (a) and corresponding double occupation increase (b) of the $L = 96$ lattice induced by 40 projectile impacts modelled by a Gaussian potential (red) as well as a Coulomb potential (green). Same input parameters as in Fig.~\ref{Fig.potentialEnergyDoubOcc}.}
\label{Fig.potentialsNs96}
\end{figure}

We now consider a larger system with $L = 96$ sites that is exposed to a further increased number of excitations, $N_{\textnormal{x}} = 40$, the results are presented in 
Fig.~\ref{Fig.potentialsNs96}.
While the overall behavior is the same as in Fig.~\ref{Fig.potentialEnergyDoubOcc}, here we observe larger deviations in $d_{\textnormal{av}}$ between the Gaussian and Coulomb models, cf. Fig.~\ref{Fig.potentialsNs96}~(b). The reason is the local excitation of a single lattice site, in case of the Gaussian model. With increase of the cluster size a growing number of sites remains unaffected, in contrasted to the long-range Coulomb model.
\\
A novel observation is the slight decrease of the average double occupation during the first few ion impacts. This is observed for both potentials,  even though the total energy show different behaviors in the two cases.
This effect will be further investigated in the following sections.

\subsection{Dependence of doublon creation on the cluster size}
\label{subsec.doublon-clustersize}
\begin{figure}[h]
\centering
	\includegraphics[scale=1]{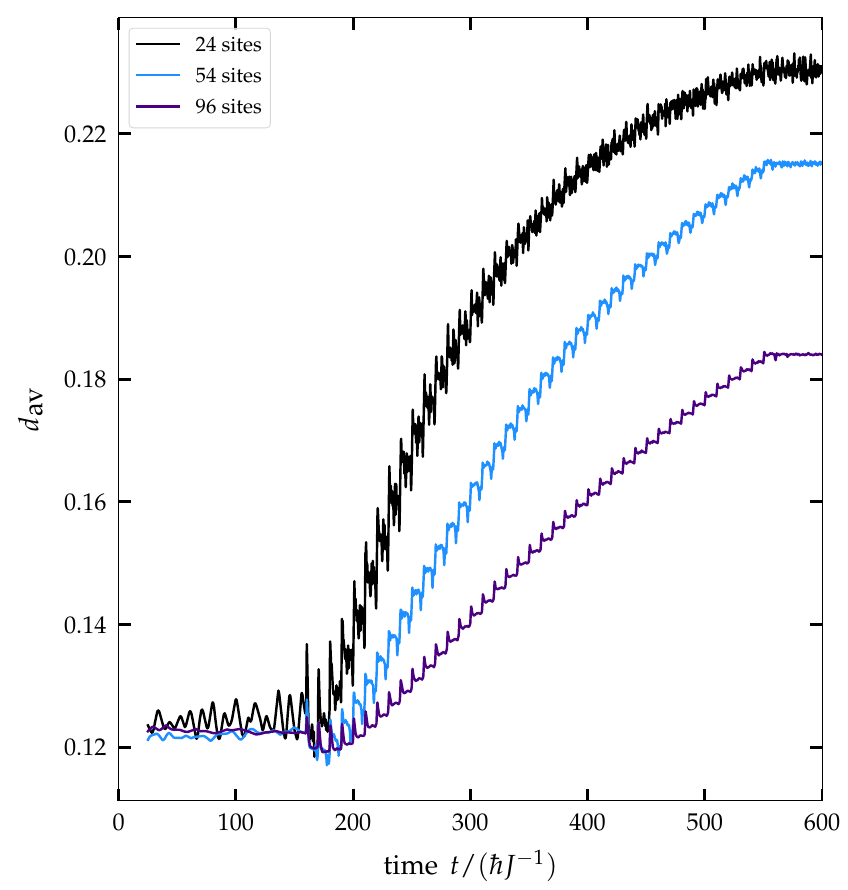}
\caption[Average double occupation for different cluster sizes]{Cluster averaged double occupation increase through $N_{\textnormal{x}} = 40$ Coulomb excitations for three honeycomb lattice sizes: $L = 24$ (black), $L = 54$ (blue) and $L = 96$ (purple). Same parameters as in Fig.~\ref{Fig.potentialEnergyDoubOcc}.}
\label{Fig.DoublonVergleich}
\end{figure}

In the following, the effect of the system size on the double occupation dynamics is investigated in more detail.
Fig.~\ref{Fig.DoublonVergleich} shows Coulomb results for lattices of three sizes, $L=24, 54, 96$. While the overall trend of an increase of the doublon number is observed for all systems, for a larger lattice the increase is slower.
The larger the lattice, the more excitations are needed to reach a specific value for $d_{\textnormal{av}}$, since the projectile energy is shared between a growing number of electrons, in agreement with the observations of Balzer \textit{et al.}~\cite{balzer_prl_18}.
Moreover, for the same reason, the peak height resulting from individual impacts and the noise in between two consecutive excitations is reduced when the lattice grows.
Another interesting observation is that the broad global minimum mentioned in the previous section seems to widen for growing cluster sizes.
Thus, more excitations are needed to escape that minimum.\\

\begin{figure}[h]
\centering
	\includegraphics[scale=0.7]{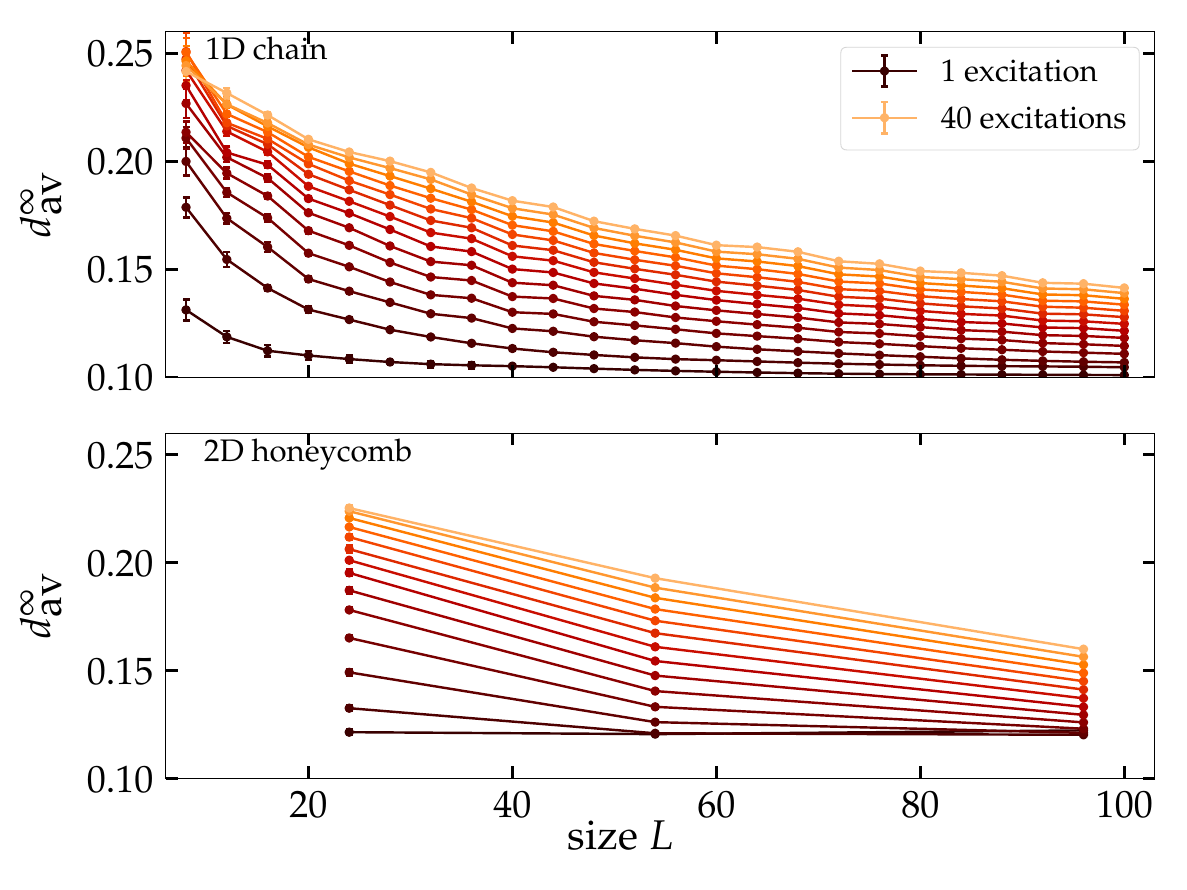}
\caption[Extension of Fig.~\ref{Fig.cascade-new} to a more realistic setup]{Asymptotic value of the mean double occupation for 1D chains consisting of up to $L = 100$ sites (top panel) and 2D honeycomb clusters with a maximum of $L = 96$ sites (bottom panel). All systems are excited between one (black curve) and 40 times (light orange) and every third excitation is plotted as a curve. The Gaussian model is used, and the initial values were chosen the same as in Ref.~\cite{balzer_prl_18}.}
\label{Fig.cascade-extension}
\end{figure}

\begin{figure}[h]
\centering
	\includegraphics[scale=1]{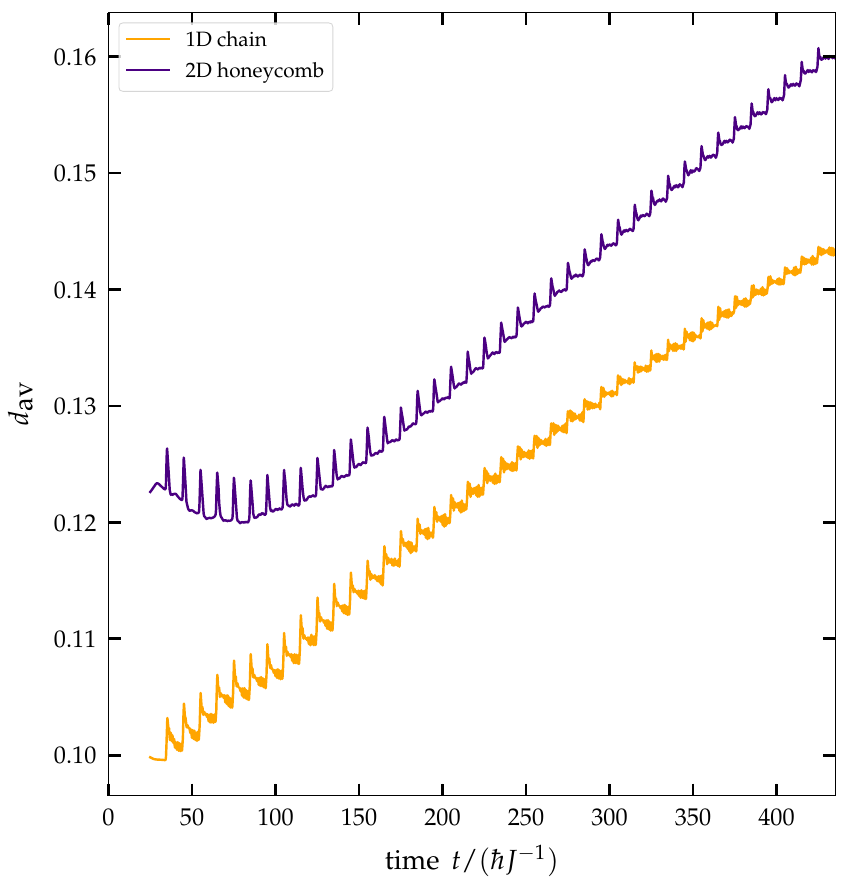}
\caption[Comparison of the average double occupation dynamics for a one- and two-dimensional system setup]{Cluster averaged double occupation dynamics for the $2$D honeycomb lattice (purple) and the $1$D chain (orange). Both systems contain $L = 96$ sites and are excited $40$ times. The Gaussian potential was used to mimic the projectile, and the input parameters match those of Fig.~\ref{Fig.cascade-extension}.}
\label{Fig.doublon1d2d}
\end{figure}

We now investigate the asymptotic value of the mean double occupation,
\begin{align}
d_{\textnormal{av}}^{\infty} = \lim_{t\to \infty} \frac{1}{\Delta t} \int_{t}^{t+\Delta t} \d\cbar{t}\, d_\tn{av}(\cbar{t})\, ,
\end{align}
which is computed after each excitation. This quantity was analyzed in Ref.~\cite{balzer_prl_18} for moderate size 1D and 2D honeycomb clusters using HF-GKBA simulations. We now apply the G1--G2 scheme which allows us to  extend the analysis to twice as large systems  and to larger numbers of excitations. To compare with Ref.~\cite{balzer_prl_18}, we employ the Gaussian potential and use the same input parameters.
Overall, the larger systems monotonically continue the trends of the previous results, which is shown in Fig.~\ref{Fig.cascade-extension}.
We also find that the  G1--G2 scheme yields very good agreement with the the previous HF-GKBA results, further comparisons have been made in Ref.~\cite{borkowski_BA_21}.
Moreover, the previously addressed broad global minimum is observable in the bottom panel at $L = 96$, as the curve of the first excitation yields a higher averaged double occupation than the following, corresponding to $N_{\textnormal{x}} = 4$.
This minimum is not visible for the one-dimensional setup.
Furthermore, we observe that the 2D setup features a much greater basic level of the averaged double occupation than the 1D one.\\
The effect of the dimensionality is explored more in detail in Fig.~\ref{Fig.doublon1d2d} where we compare 1D and 2D clusters containing the same number of sites, $L=96$.
The figure confirms that the broad global minimum of $d_\tn{av}({t})$ does not appear in a linear chain (orange curve), but is restricted to higher dimensionality. 
Furthermore, the initial double occupation of the honeycomb lattice is significantly larger than the corresponding value of the chain setup and the increase of $d_\tn{av}({t})$ is faster.
This is explained by the increased number of nearest neighbors of a lattice site which supports the build-up of correlations.
%
\subsection{Dependence of doublon creation on the \\time interval between impacts}
\label{subsec.FreqVar}
\begin{figure}[h]
\centering
	\includegraphics[scale=1]{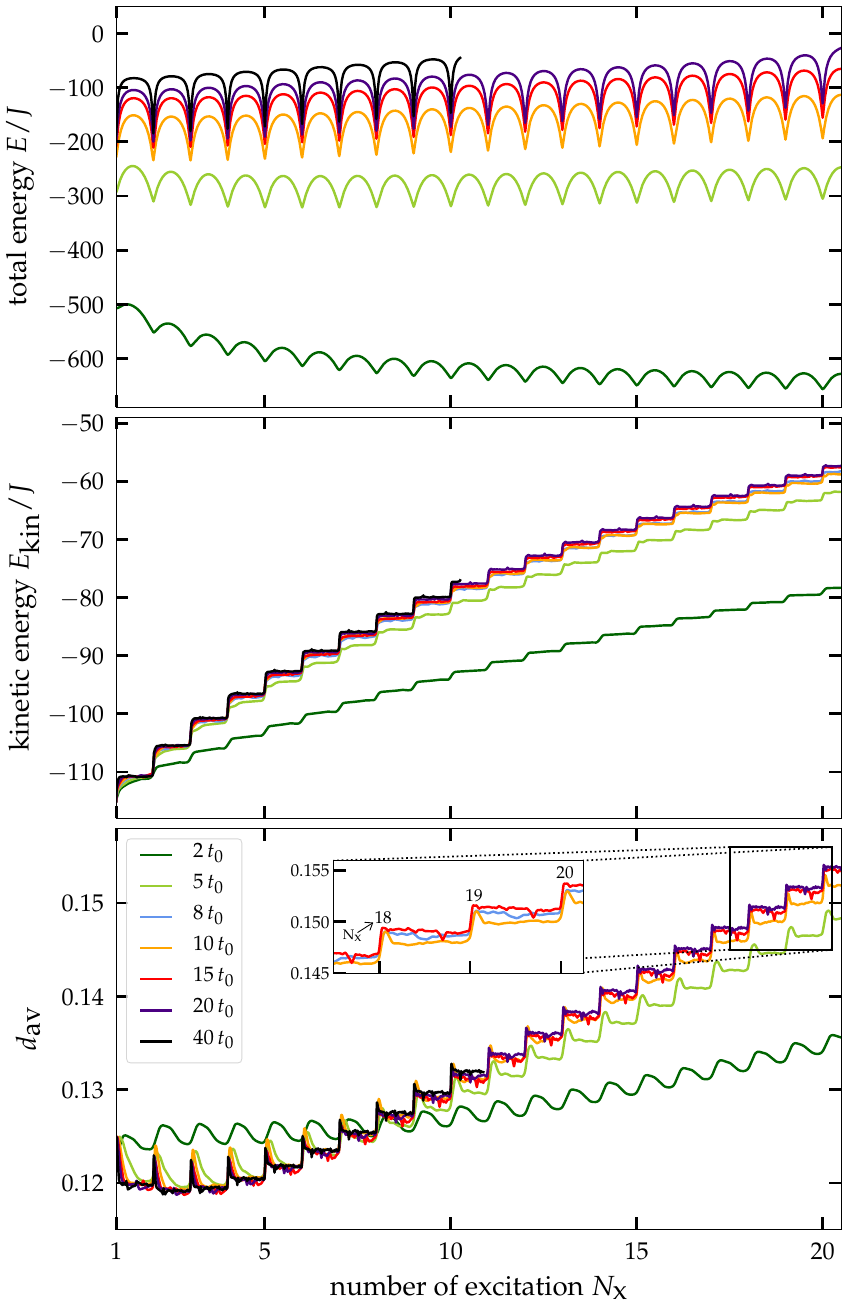}
\caption[Variations in impact frequency for a lattice with $L = 96$ sites]{Dependence of total energy $E$ (top figure) and average double occupation $d_{\textnormal{av}}$ (bottom figure) on the number of excitations $N_x$ and the time interval between them, for a $96$ sites  honeycomb cluster. The six different colors denote different time intervals, $\Delta t_{\textnormal{imp}}$, in between two consecutive excitations. The orange line ($10 \, t_0$) corresponds to the interval used in all  calculations above. 
The initial position and velocities of the projectiles match those introduced in Sec.~\ref{subsec.GaussCoulomb}.}
\label{Fig.differentFrequencies}
\end{figure}

\begin{figure}[h]
\centering
	\includegraphics[scale=1]{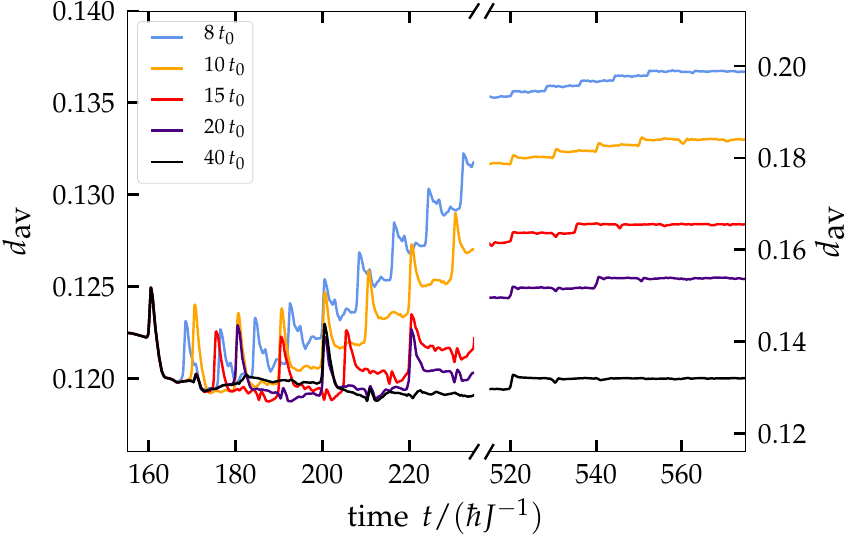}
\caption[Variations in impact frequency -- further results]{Cluster averaged double occupation, as shown in Fig.~\ref{Fig.differentFrequencies}, but now plotted versus time. In the right part, the final stage of the time evolution around the final impact is displayed (note the brake of the x-axis).}
\label{Fig.differentFreqSeveralPlots}
\end{figure}

We now vary the frequency of ion impacts in a broad range. 
The results are shown in Fig.~\ref{Fig.differentFrequencies} for the largest honeycomb cluster ($L = 96$) using the Coulomb potential.
The upper panel illustrates how the total energy of the lattice evolves with the number of excitations for the respective impact frequencies.
All calculations have been performed for the same total time duration, $t_\tn{max} = 650 \, t_0$.
For this reason, the total energy is much lower for high impact frequencies (green), since there are up to twenty times as many excitations compared to the low frequency case, the potentials of which superpose each other.
The flattening of the ion-impact peaks in the energy can also be attributed to that superposition, since the interim energy loss of the system directly induced through an impact becomes less relevant with every additional long-range Coulomb potential applied to the lattice electrons.
To see the direct influence of the frequency variation, we consider the more sensitive kinetic energy [cf. Eq.~\eqref{eq:kinetic_energy}] in the middle panel of Fig.~\ref{Fig.differentFrequencies}. Here we see a similar build-up behavior for most time intervals $\Delta t_\tn{imp}$, except for the highest ion frequencies, where the energy growth is hampered. For these cases, the ion frequency is on a similar time scale to the immediate electron dynamics in the target. Therefore, the finite correlation spreading and carrier mobility prevent a faster energy transfer.\\
This effect becomes even clearer for the double occupation (cf. bottom panel of Fig.~\ref{Fig.differentFrequencies}). Again, we observe that for sufficiently large $\Delta t_\tn{imp}$, the successive double-occupation build-ups coincide whereas higher frequencies result in a decreased influence per projectile.  
A peculiarity is found for $\Delta t_\tn{imp} = 10t_0$ (orange curve), for which $d_\tn{av}$ lies slightly below the respective curves for $\Delta t_\tn{imp} = 8t_0$ (blue) and $\Delta t_\tn{imp} = 15t_0$ (red), see the inset of Fig.~\ref{Fig.differentFrequencies}.
This non-monotonic behavior can be attributed to an interference of the impact frequency with one of the systems characteristic frequencies
and will be further explained in context of Fig.~\ref{Fig.differentFreqSeveralPlots}.

Another interesting observation is that the broad global minimum of $d_\tn{av}$ vanishes, for very high frequencies. This indicates that the effect is caused by electronic correlation effects on longer time scales.
To further investigate the non-monotonic effect in the double occupation, we now focus on the noise in between two excitations and the vibration of $d_{\textnormal{av}}$.
In Fig.~\ref{Fig.differentFreqSeveralPlots}, the time evolution of the average double occupation is shown for different impact frequencies.
Obviously, higher frequencies lead to a faster increase of the double occupation, due to the larger number of excitations. Note that, there is a feature that is observed for all impact frequencies: a small kink in $d_{\textnormal{av}}$ (upwards for early times and downwards, for later times) appears at approximately $10 \, t_0$ after each projectile impact. 
This indicates that there exists a characteristic frequency in the $L=96$ honeycomb cluster.
For $\Delta t_{\textnormal{imp}} = 10 \, t_0$ (orange), this frequency coincides with the one of the ion impacts.
This explains the unusual behavior of the double occupation for this case: instead of the small local minima, here maxima are observed.

To find an explanation for the mentioned characteristic frequency of the system, we now investigate the space-resolved carrier redistribution in the lattice.

\subsection{Space-resolved electron dynamics}\label{ss:space-resolved}
\begin{figure}[h]
\centering
	\includegraphics[scale=1]{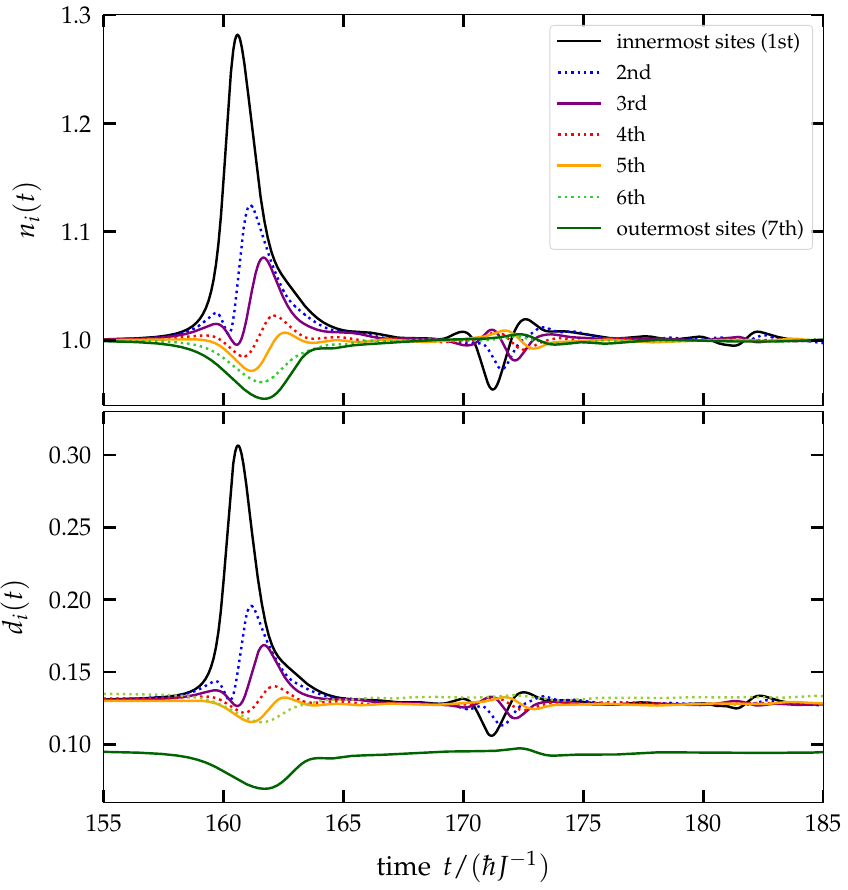}
\caption[site-resolved density and double occupation dynamics]{Time evolution of the local density (top) and double occupation (bottom) on the different honeycomb rings for an ion impact in the center. Due to the highly symmetric setup, the results coincide for all lattice sites of a single ring.}
\label{Fig.siteResolvedCentralImpact}
\end{figure}

The highly symmetric $96$-site honeycomb cluster consists of four layers of hexagons. To increase the space resolution we consider, instead, seven rings of sites, 
(sites on the $n$th ring are those that can be reached from the innermost ring within $n-1$ steps), 
cf. Fig.~\ref{Fig.lattice}. In Fig.~\ref{Fig.siteResolvedCentralImpact}, we show the time evolution of the average electron density (top) and double occupation (bottom) within each of these rings, following a single ion excitation through the center of the system.
%
For the density, we see a drastic initial increase on the innermost ring that follows from the direct Coulomb attraction by the charged projectile. Simultaneously, the outer rings are slightly depopulated as the electrons move towards the center. Subsequently, we observe a propagating density wave with a finite velocity through the honeycomb rings. For the outer rings, the density change becomes less pronounced, as the electrons spread across a larger number of lattice sites. After a short relaxation phase, the local densities exhibit a second peak (with lower amplitude and opposite sign), although the ion is already far away from the target and no additional excitation is invoked. Thus, the oscillation revival is caused by an intrinsic property of the finite cluster. The time interval between both events is approximately $\Delta t = 10 t_0$, which perfectly agrees with the observed time delay of the energy kinks in Fig.~\ref{Fig.differentFreqSeveralPlots}. 
Note that a (less pronounced) second revival is observed after additional $10 t_0$ (cf. the black curve in Fig.~\ref{Fig.siteResolvedCentralImpact}). \\
The double occupation (bottom of Fig.~\ref{Fig.siteResolvedCentralImpact}) exhibits a similar behavior as the density which is explained by the mean field contribution to the doublon number. 
Note that for the outermost sites, we see a significantly lower double occupation---a property that is already present in the ground state and at the initial time, due to the reduced connectivity on the edges of the system. 

\subsection{Localized versus random impacts}\label{ss:impact-point}
\begin{figure*}[t]
\centering
	\includegraphics[width=\textwidth]{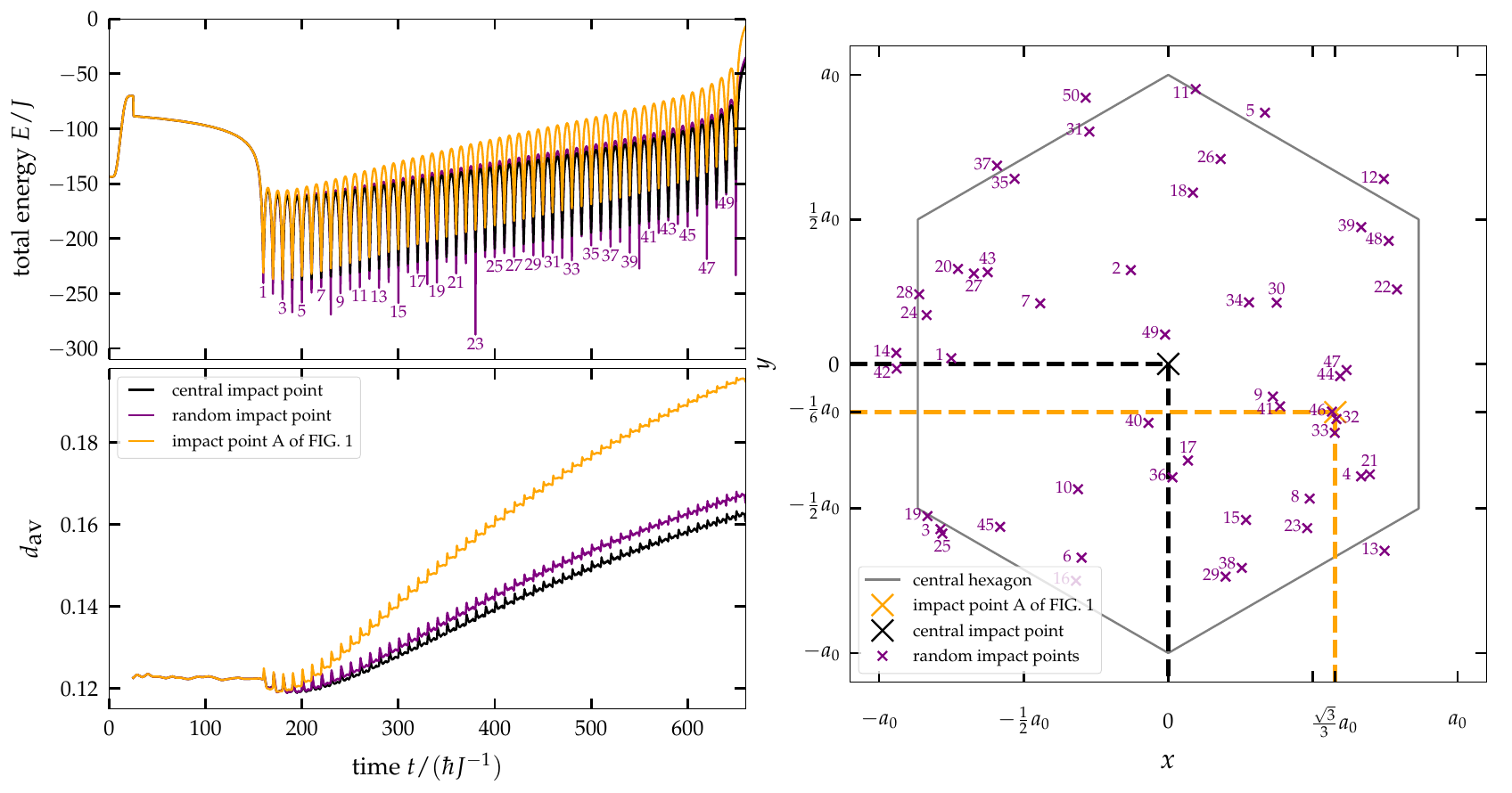}
\caption[Randomized impact points for $L = 96$]{Response of a 96 site lattice to 50 excitations for three different selections of impact points (sketched in the right figure): black line: central impact; yellow: fixed impact point as in Fig.~\ref{Fig.lattice}; purple: random impact points, uniformly distributed within a radius of $1a_0$.
\textbf{Top left}: time evolution of total energy of the lattice $E$ and \textbf{bottom left}: time evolution of average double occupation $d_{\textnormal{av}}$ -- the realized 50 impact points shown in the right panel are numbered in chronological order.
The projectiles' initial velocity is the same as in Fig.~\ref{Fig.potentialEnergyDoubOcc}.}
\label{Fig.randomImpact}
\end{figure*}

We now consider a different setup where the impact point of the projectiles is varied.
To this end, we generate uniformly distributed random positions within a circle of radius $1a_0$. For a fixed time interval of $\Delta t_{\textnormal{imp}}=10t_0$, we simulate successive ions of the same energy (cf. Eq.~\eqref{eqn.yukawaCoulombInput}) penetrating through those randomized impact points. All computations are performed with a number of $N_{\textnormal{x}} = 50$ excitations.
The results are presented in Fig.~\ref{Fig.randomImpact} for
the total energy of the lattice (top), the cluster-averaged double occupation (bottom) and the impact positions (right).
As a guide to the eye, the innermost hexagon of the $96$-site cluster is shown (grey lines). The impact-point distribution is illustrated (purple) alongside of point A in Fig.~\ref{Fig.lattice} (orange) and the central position (black). The purple numbers indicate the order of the excitations.
From the behavior of the lattice energy and of the double occupation, we conclude that the random impact scenario leads to results that are very similar to the previous findings. In fact, both quantities are predominantly enclosed between the curves for the impact point A (orange) and the central impact (black). Thus, we conclude that the spatially uniform distribution of ion impacts does not significantly change the electronic behavior. A peculiar feature is the random fluctuation of the energy minima which is caused by the varying distance of the projectile to the nearest lattice sites. At the same time, the average double occupation is not affected by these variations and exhibits the same monotonic trend as for the central impact point (black curve). Furthermore, we recover the broad global minimum, discussed in Sec.~\ref{subsec.doublon-clustersize}. 




\section{\label{sec.discussion}Conclusions}
In summary, we have extended the analysis of electronic correlations, in particular, doublon formation, in small hexagonal graphene-type clusters  caused by ion impact, that was presented in a recent Letter \cite{balzer_prl_18}. There the scenario of multiple ions hitting the target in the same spot and at fixed time intervals was analyzed, using a nonequilibrium Green functions analysis coupled to an Ehrenfest treatment. The large computational effort of NEGF simulations limited the parameter range that could be studied.

Here, we applied the recently derived G1--G2 scheme \cite{schluenzen_20_prl,joost_prb_20} to this problem. The advantageous linear scaling with the simulation duration of this approach has allowed us to significantly extend the previous analysis: to larger clusters, to more impacts, as well as to a variation of the impact point and the time interval between impacts. We confirmed that increasing the number of impacts allows to further increase the mean doublon number in the cluster until it eventually approaches the mean field limit $0.25$. When the system size is increased the increase of the doublon number slows down since the impact energy is shared among a growing number of electrons. An interesting observation was that, in a 2D hexagonal arrangement of $L$ lattice sites, the average doublon number increases significantly faster compared to a linear arrangement of the sites.

Our fully time-dependent
approach is able to resolve non-adiabatic processes in the electronic sub-systems and effects beyond linear response. Non-adiabatic effects are particularly evident at small time intervals between subsequent impacts because the electronic system has not enough time to return to the ground state before the next impact occurs.
This can be clearly seen in the behavior of the average double occupation of the electrons shown in 
Fig.~\ref{Fig.differentFrequencies}. While for large time intervals between projectiles the first few impacts result in a global minimum in the double occupation, the situation changes when the time interval is being reduced. For the interval $5t_0$ only after reaching the global minimum subsequent impacts lead to a periodic but steady increase in the number of doublons, for $2t_0$, this behavior is present from the very first impact and no such minimum can be observed.


We analyzed two protocols of ion impact that can be realized in various experimental setups. First, we studied a strictly periodic sequence of ions that hit the same lattice site; this can be realized with ion guns. Second, we studied a situation where the impact point of the ions is chosen randomly. This case is closer to a gas or a plasma. Interestingly, we observed that the result for the average doublon number is almost the same as in the case of a fixed impact point. What could be done next is to randomly sample the time interval between ions as well in order to properly reflect the ion velocity distribution in the gas or plasma. This will allow to make quantitative predictions for the interaction of plasmas with solids, e.g. Ref.~\cite{Bonitz_fcse_19}.  

Additional questions that are of interest in the context of ion-solid interaction are innerionic processes such as ion neutralization \cite{balzer_cpp_21}, electronic excitations and secondary electron emission, e.g. Ref.~\cite{Pamperin_Bronold_PRB_2015}. Such processes have recently  become accessible to accurate measurements for highly charged projectiles impacting graphene and other correlated 2D materials, e.g. Refs.~\cite{wilhem_prl_17,niggas_comm_phys_21},  
and are still missing a full theoretical description. The present NEGF approach within the G1--G2 scheme provides the proper starting point for this problem as it allows to resolve the full dynamics of electronic correlations. At the same time, the present Hubbard model needs to be extended to include additional bands~\cite{konova_acs.nanolett_21}.


\acknowledgments

We thank Karsten Balzer for helpful discussions.

\end{document}